\documentclass[twocolumn,showpacs,superscriptaddress,floatfix,prl]{revtex4-1}
\usepackage{graphicx,amsfonts,amssymb,amsmath,hyperref,hypcap,enumerate}

\usepackage{amsthm}
\usepackage{multirow}
\usepackage{graphicx}
\usepackage{dcolumn}   
\usepackage{bm}        
\usepackage{amssymb}   
\usepackage{amsmath}
\usepackage{epstopdf}
\usepackage{color}

\DeclareMathAlphabet{\mathitb}{OT1}{cmr}{bx}{sl}

\begin{document}
	\title{Unconventional scaling theory in disorder-driven quantum phase transition}
\author{Xunlong Luo}
\affiliation{International Center for Quantum Materials, Peking University, Beijing 100871, China}
\affiliation{Collaborative Innovation Center of Quantum Matter, Beijing 100871, China}
\author{Tomi Ohtsuki}
\affiliation{Department of Physics, Sophia University, Chiyoda-ku, Tokyo 102-8554, Japan}
\author{Ryuichi Shindou}
\email{rshindou@pku.edu.cn}
\affiliation{International Center for Quantum Materials, Peking University, Beijing 100871, China}
\affiliation{Collaborative Innovation Center of Quantum Matter, Beijing 100871, China}

\date{\today}
	\begin{abstract}
We clarify novel forms of scaling functions of conductance, 
critical conductance distribution and localization length in a disorder-driven quantum phase 
transition between band insulator and Weyl semimetal phases. 
Quantum criticality of the phase transition is controlled by a clean-limit 
fixed point with spatially anisotropic scale invariance. 
We argue that the anisotropic scale invariance is reflected on unconventional scaling function forms 
in the quantum phase transition. We verify the proposed scaling function forms in 
terms of transfer-matrix calculations of conductance and localization length in a tight-binding model.  
	\end{abstract}
\maketitle

Scaling theories play a central role in the studies of Anderson localization~\cite{anderson58,wegner76} as well as 
other disorder-driven quantum phase transitions. Inspired by the finite size scaling theory
by the gang of four~\cite{abrahams79}, scaling theories of localization 
length~\cite{mackinnon81,pichard81}, and conductance~\cite{anderson80,slevin01prl} 
have been developed and become the core of our current understandings of the 
localization phenomena. The theories facilitate numerical  
studies of the phenomena, that establish 
a rich variety of the universality classes~\cite{wigner51,dyson62a,dyson62b,zirnbauer96,altland97}. 
All the Wigner-Dyson universality classes 
are characterized and distinguished from one another by critical and dynamical 
exponents, and critical conductance distribution (CCD)~\cite{shapiro90,cohen92,slevin97,slevin00}. 
Meanwhile, all of them obey the similar scaling functions;
\begin{eqnarray}
Q(L,m,\Delta_1,\Delta_2,\cdots) = F_{Q}(m^{\nu} L,m^{|y_1|}\Delta_{1},\cdots). \label{conventional}
\end{eqnarray}
Here $Q$ is a (properly normalized) dimensionless physical quantity, $L$ is a linear dimension of the system 
size, $m$ is a relevant scaling variable with its scaling dimension $\nu$, and $\Delta_{j}$ ($j=1,2,\cdots$) 
is an irrelevant scaling variable with negative scaling dimension $y_{j}$. 
Naturally one may raise a question by asking ``Is there any new disorder-driven quantum phase 
transition that obeys different {\it forms} of scaling functions ?'' 

In this rapid communication, we answer this question affirmatively, by investigating quantum criticality of a 
disorder-driven phase transition between band insulator (BI) and Weyl semimetal (WSM) phases. 
We clarify novel forms of scaling functions of conductance, CCD and localization length such 
as in Eqs.~(\ref{conductance-scaling-2}), (\ref{xix-scaling-2}), (\ref{xiz-scaling-2}), 
(\ref{xiz-scaling-3}), and (\ref{xix-scaling-3}). 
The criticality of the BI-WSM transition is controlled by a fixed point 
in the clean limit that has spatially anisotropic scale invariant 
property~\cite{cardy,yang1,carpentier,yang2,jrwang,roy16arXiv,luo18}. 
We show that the anisotropic scale invariance results 
in unconventional forms of scaling functions for conductance, CCD and 
localization length in the disorder-driven BI-WSM quantum 
phase transition. Based on numerical simulations on 
a lattice model with disorders, we demonstrate the validity of the 
proposed scaling properties. 

\begin{figure}[t]
\centering
\includegraphics[width=0.8\linewidth]{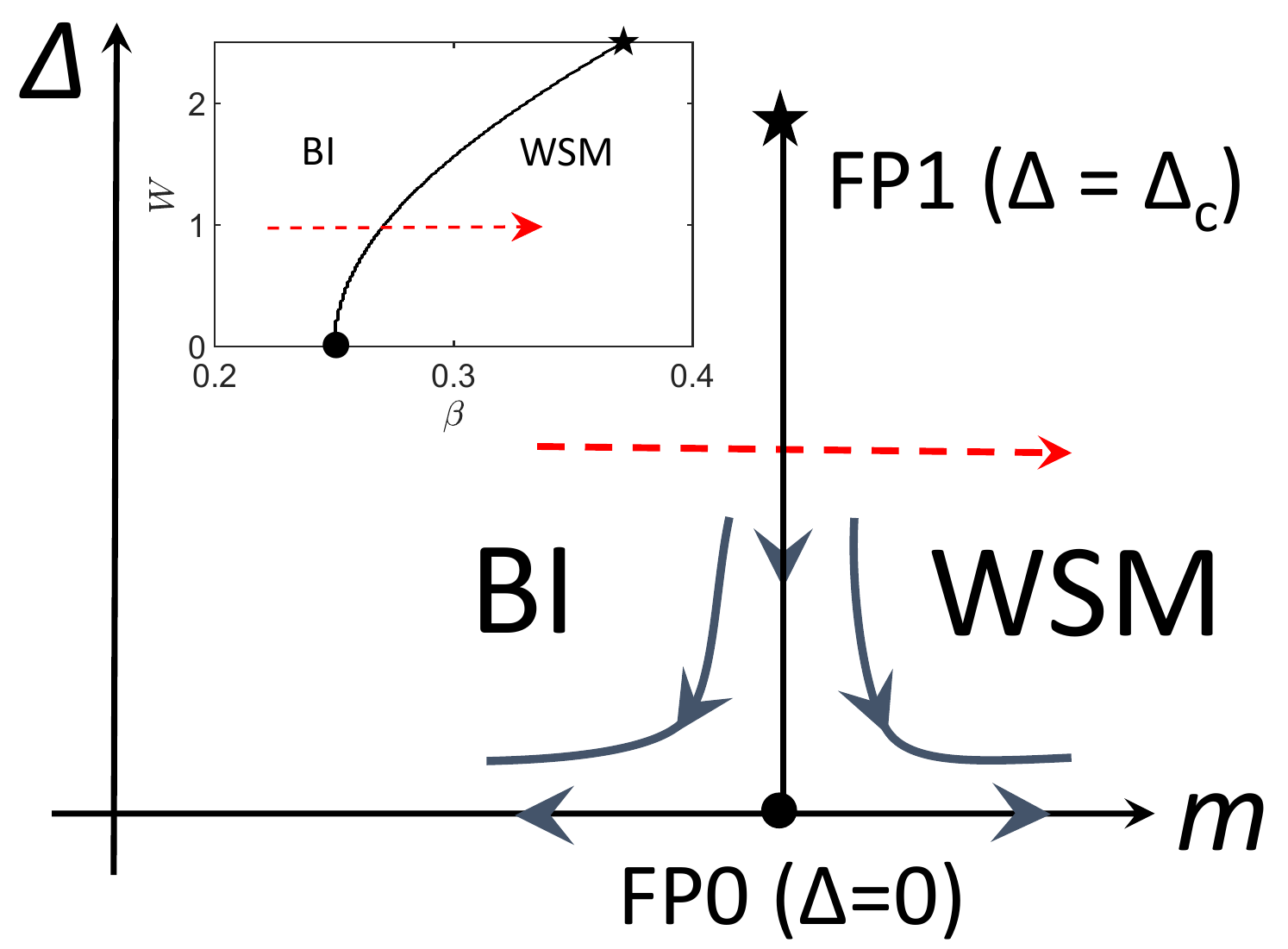}
\caption{(color online) Schematic of the  
quantum phase transition between BI and WSM phases. The dark blue 
arrows denote renormalization group (RG) flows~\cite{roy16arXiv,luo18}.  The criticality of 
the quantum phase transition at finite disorder strength is 
controlled by the critical point in the clean limit (`FP0' denoted by 
$\bullet$ mark). For the stronger disorder strength side, the quantum phase 
transition line is terminated by a quantum critical 
point (`FP1' denoted by $\star$ mark). Inset: a phase diagram 
of the tight-binding model for a three-dimensional layered Chern 
insulator with disorders~\cite{cz-chen15,s-liu16,roy16arXiv,luo18}. 
The disorder strength $W$ and interlayer 
coupling strength $\beta$ correspond to $\Delta$ and the effective 
mass $m$ respectively. The disorder strength $W$ as well as the 
interlayer coupling strength $\beta$ drives the quantum 
phase transition between BI (3D Chern insulator) and WSM phases.
In Figs.~\ref{fig:2+3} and \ref{fig:8}, 
we change the effective mass $m$ (interlayer coupling 
strength $\beta$) with fixed $\Delta$ (disorder strength $W$); dashed red 
lines with arrow. In Fig.~\ref{fig:5}, the system 
is on the BI-WSM phase transition line. In the lower panel of 
Fig.~\ref{fig:5}, the system is on (or very close to) the FP1.}
\label{fig:1}
\end{figure}

Weyl semimetal (WSM) is a class of three-dimensional semimetal that has a band touching point 
with linear dispersions along all the three 
directions (`Weyl node')~\cite{murakami07,balents11,weng15,sy-xu15,bq-lv15}. The Nielsen-Ninomiya 
theorem dictates that two band touching points with the linear dispersions must appear 
in a pair in the first Brillouin zone. When a pair of two Weyl 
nodes annihilate with each other, the system undergoes a 
quantum phase transition from WSM to BI phases. The phase transition is described 
by an effective continuum model with a magnetic dipole in the momentum space~\cite{yang1,yang2,roy16arXiv,luo18},  
\begin{align}
{\cal H}_{\rm eff} &= \int d^2{\bm x}_{\perp} dz \!\ \psi^{\dagger}({\bm x}) 
\Big\{ iv \big(\partial_{x} {\bm \sigma}_x 
+ \partial_y {\bm \sigma}_y \big) \nonumber \\
&\ \ \ - \big((-i)^2 b_2 \partial^2_z - m \big){\bm \sigma}_z  
\Big\} \psi({\bm x}), \label{eff0}  
\end{align} 
with ${\bm x}_{\perp} \equiv (x,y)$ and ${\bm x}\equiv 
({\bm x}_{\perp},z)$. ${\bm \sigma}_{\mu}$ ($\mu=x,y,z$) are 
2 $\times$ 2 Pauli matrices. For positive $m$ (WSM phase), the electronic 
system at $E=0$ has a pair of Weyl nodes with the opposite magnetic monopole 
charges at ${\bm k}_{\rm MM}=(0,0,\pm\sqrt{m/b_2})$; 
`magnetic dipole' in the momentum space.  For negative $m$ 
(BI phase), the system has an energy gap at the zero energy. Previously, the stability 
of the critical point ($m=0$) against the Coulomb interaction~\cite{yang1,yang2} 
as well as short-ranged disorder~\cite{yang1,roy16arXiv,luo18} has been studied.  
Especially, a tree-level renormalization group analysis on the continuum model dictates 
that the quantum critical point at $m=0$ is robust against any types of short-ranged 
disorder~\cite{yang1,yang2,roy16arXiv,luo18,fradkin86,goswami11,syzranov15prb,syzranov15prl}. 
Thus, small but finite disorder is always renormalized to the critical point 
in the clean limit, as long as the disorder strength is smaller than a certain critical value 
$\Delta_c$ (Fig.~\ref{fig:1}). Quantum criticality of the disorder-driven 
BI-WSM quantum phase transition at finite $\Delta<\Delta_c$ is controlled by 
the clean-limit fixed point at $m=\Delta=0$. 
We dub the fixed point as `FP0' as in Fig.~\ref{fig:1}. 

The gapless theory at $m=\Delta=0$ has a quadratic dispersion 
along the dipole ($z$) direction, while it has linear dispersions within 
the perpendicular ($xy$) directions. Thereby, the clean-limit fixed point has 
the following spatially anisotropic scale invariant property;   
\begin{align}
z^{\prime} &= b^{\frac{1}{2}} \!\ z, \ \    
{\bm x}^{\prime}_{\perp} = b \!\ {\bm x}_{\perp}, \label{scaling-zxy} 
\end{align}
with time $t^{\prime} = b \!\ t$ and the single-particle energy $E^{\prime} = b^{-1}E$. 
Hereafter a symbol for the scale change, $b \equiv e^{-dl}<1$, counts how many 
times we carry out a renormalization. Quantities with 
and without prime denote those after and before the renormalization,  
respectively. 
As we will see below, the anisotropic scaling leads to 
new forms of scaling functions for the conductance 
and localization length at the Weyl nodes ($E=0$). 

Let us begin with the scaling property of the zero-energy conductance.  
According to the anisotropic scaling, the density state per volume $\rho(E)$ scales 
as $\rho^{\prime}(E^{\prime})=
b^{-(d-\frac{1}{2}-1)}\rho(E)$ at the fixed 
point (FP0)~\cite{yang1,roy16arXiv,luo18}, 
where $b^{-(d-\frac{1}{2})}$ and $b^{+1}$ come from 
Eq.~(\ref{scaling-zxy}) and $E^{\prime} = b^{-1}E$, respectively. 
The diffusion constant along the 
dipole direction scales as $D^{\prime}_{z}=b^{-(1-1)}D_z=D_z$, while that along 
the perpendicular directions scales as 
$D^{\prime}_{\perp}=b^{-(1-2)}D_{\perp}$~\cite{luo18}. 
Thus, the Einstein relation, $\sigma_{\mu}\equiv e^2 D_{\mu} \rho$, gives the 
conductivity scaling at the fixed point in the clean limit as 
$\sigma^{\prime}_{z}=b^{-d+\frac{3}{2}} \sigma_z$ and 
$\sigma^{\prime}_{\perp}=b^{-d+\frac{5}{2}} \sigma_{\perp}$ respectively~\cite{luo18}. 
With $G_{z}\equiv \sigma_z L^{d-1}_{\perp}/L_z$ and 
$G_{\perp} \equiv \sigma_{\perp} L^{d-3}_{\perp} L_z$, one naturally 
reaches the following scaling relations of the zero-energy conductances 
under the renormalization;
\begin{align}
G^{\prime}_{\mu}(L^{\prime}_{z},L^{\prime}_{\perp},\Delta^{\prime},m^{\prime}) 
= G_{\mu}(L_z,L_{\perp},\Delta,m), \label{conductance-scaling}
\end{align}
with $\mu=\perp,z$, $L^{\prime}_z=b^{\frac{1}{2}}L_{z}$, $L^{\prime}_{\perp}=bL_{\perp}$, 
$\Delta^{\prime}=b^{-y_{\Delta}} \Delta$, and $m^{\prime}=b^{-1}m$. $y_{\Delta}$ 
is a scaling dimension of the short-ranged disorder strength $\Delta$ and 
is negative, $y_{\Delta} = -d+\frac{5}{2}<0$ ($d=3$). 
 
Eq.~(\ref{conductance-scaling}) generates all the scaling properties of the zero-energy 
conductances near the BI-WSM phase transition. 
We start with tiny $m$ and renormalize many times 
until the relevant scaling variable $m$ goes far away from the critical point, say 
$m^{\prime}=1$. Solving $b$ in favor for small $m$, we obtain a scaling function 
of the conductances as, 
\begin{align}
G_{\mu}(L_z,L_{\perp},\Delta,m) = \Phi_{\mu}(m^{\frac{1}{2}} L_z, mL_{\perp}, 
m^{|y_{\Delta}|} \Delta). \label{conductance-scaling-2}
\end{align} 
For smaller $m$, we may replace the third argument by zero. The conductance 
scaling function depends on the linear dimension of system size along the 
dipole direction and that along the perpendicular directions with different 
exponents in $m$. This unconventional scaling form comes from the 
spatially anisotropic scale invariant property at the clean-limit fixed point. 

To test this scaling function in numerical simulations, we take a 
tetragonal geometry, $L_x=L_y=L_{\perp}=\eta L^2_z$ with 
fixed geometric parameter $\eta$, to reduce Eq.~(\ref{conductance-scaling-2}) into 
a single parameter scaling form,  
\begin{align}
G_{\mu}(L_z,L_{\perp}=\eta L^2_z,\Delta,m) = \phi_{\mu}(m^{\frac{1}{2}}L_z;\eta). 
\label{conductance-scaling-3}
\end{align}
Using the same tetragonal geometry, we numerically calculate the 
conductances of a tight-binding model for a layered Chern insulator
with disorders~\cite{cz-chen15,s-liu16,roy16arXiv,luo18,supplemental}. In the tight-binding model, 
we fix a disorder strength $W$ and change an interlayer coupling strength $\beta$ (inset of Fig.~\ref{fig:1}). 
When the coupling strength exceeds a critical value $\beta_c$, the electronic system 
undergoes the quantum phase transition from BI phase ($\beta<\beta_c$) 
to WSM phase with a pair of Weyl nodes ($\beta>\beta_c$). 
The same quantum phase transition can be induced by a change 
of the disorder strength $W$ with constant $\beta$. Criticality of the quantum 
phase transition is controlled by the gapless theory in the clean limit, Eq.~(\ref{eff0}), 
where $\delta \beta=\beta-\beta_c$ is proportional to the effective mass $m$ 
in Eq.~(\ref{eff0}). In the WSM phase, the pair of the Weyl nodes 
appear at ${\bm k}_{\rm MM}=(0,0,\pm k_{z,c})$, where $k_{z,c}\propto \sqrt{\delta \beta}$. 
In the finite-size tight-binding model calculation, we choose 
$\eta=1/25$ and $(L_z,L_{\perp})=(18,13)$, $(20,16)$, 
$(24,23)$, $(26,27)$, $(30,36)$, $(32, 41)$, all of which satisfy 
$L_z = \eta L^2_z$ approximately. The conductance along the $\mu$ 
direction $G_{\mu}$ is calculated by the transfer matrix method with the 
periodic boundary conditions for the transverse directions. 
For $G_x (G_z)$, we take 40 (5000) samples to obtain their 
distribution functions. 

\begin{figure}[t]
	\centering
	\includegraphics[width=0.9\linewidth]{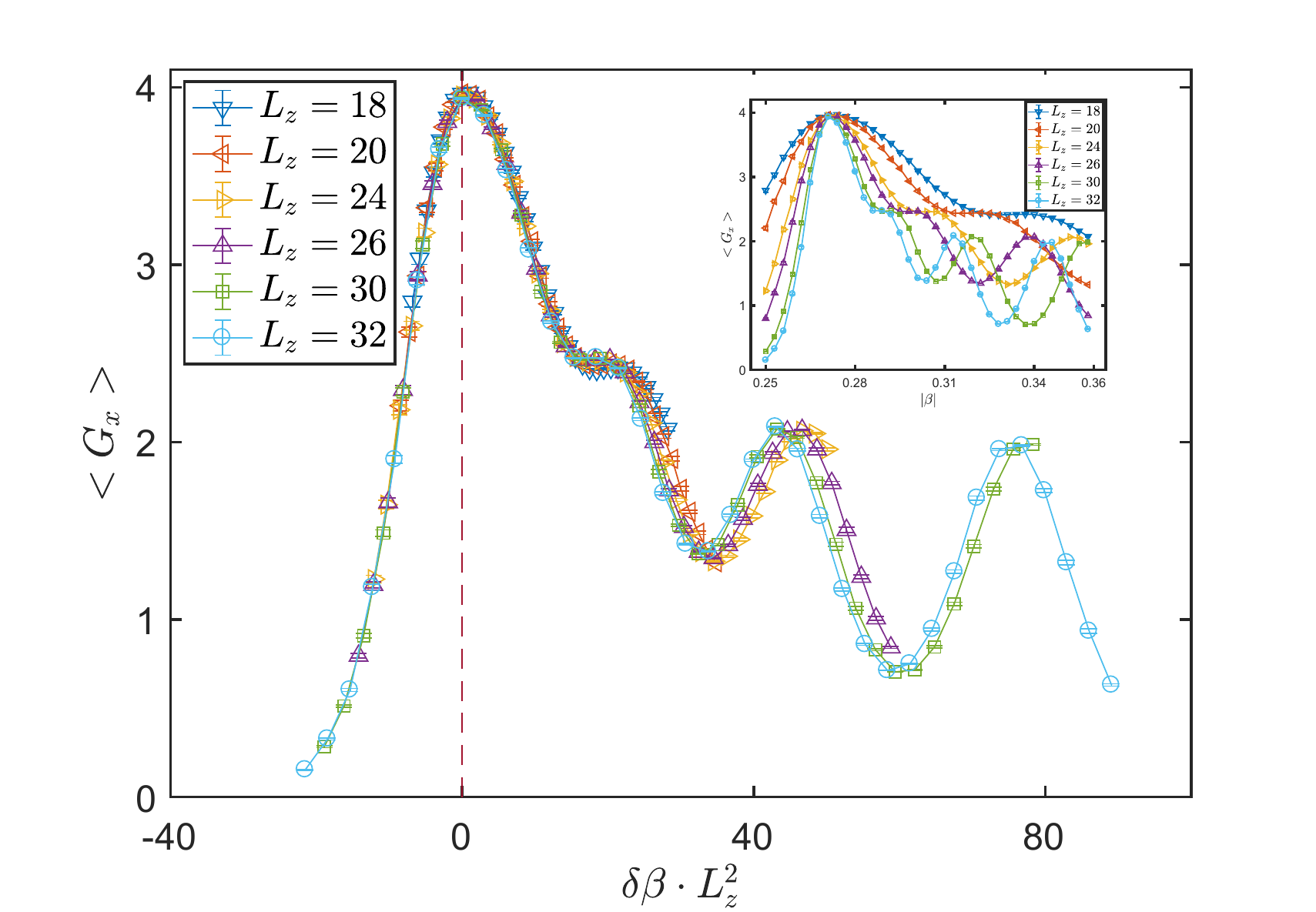}
	\includegraphics[width=0.9\linewidth]{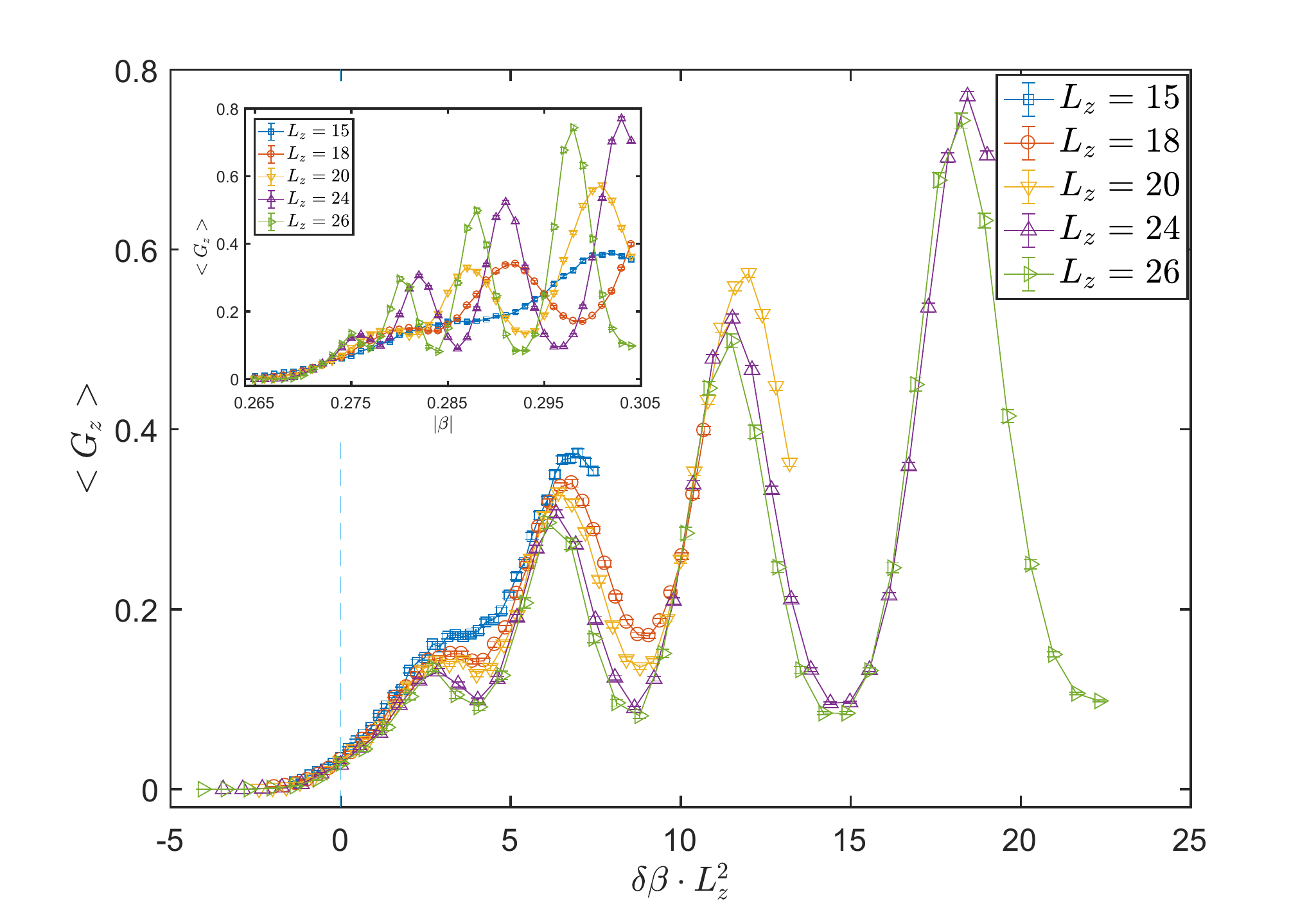}
	\caption{(color online) 
$G_x$ (upper) and $G_z$ (lower)  as a function of 
$\delta \beta \cdot L_z^{2}$ with the tetragonal geometry 
($L_{\perp}=\eta L^2_z$ with $\eta=1/25$) and different $L_z$. 
We take a set of parameters in the tight-binding model in Refs.~\cite{s-liu16,supplemental}  
as $(W,\beta)=(1,\beta_c+\delta \beta)$ with $\beta_c=0.271$. The disorder strength $W$ 
corresponds to $\Delta$ in Fig.~\ref{fig:1} and 
Eq.~(\ref{conductance-scaling-3}). $\delta \beta$ is proportional 
to the effective mass $m$ in Fig.~\ref{fig:1} and 
Eqs.~(\ref{eff0}) and (\ref{conductance-scaling-3}). Insets: $G_x$ (upper) and $G_z$ (lower) 
as a function of $\delta \beta$ with different $L_z$.}
	\label{fig:2+3}
\end{figure}

Fig.~\ref{fig:2+3} shows $G_x$ and $G_z$ as a function of 
$\delta \beta \cdot\ L^2_z$ for the constant $W$. 
Almost all the numerical data fit in the proposed novel single 
parameter scaling form, Eq.~(\ref{conductance-scaling-3}).
Especially, the data with larger system sizes near the critical point 
collapse into the form better, indicating the validity of the 
single parameter scaling form.  The conductances in the WSM phase 
side show oscillatory behaviors as a function of $\delta \beta L^2_z$. 
In the WSM phase, the Weyl points appear at ${\bm k}_{\rm MM}=(0,0,\pm k_{z,c})$. 
The finite-size system with the periodic (fixed) boundary condition can feel   
these Weyl nodes only when $k_{z,c}$ becomes equal to $2\pi/L_z$ ($\pi/L_z$) 
times integer. Thus, the conductances show peaks when $L_z k_{z,c}$ matches 
integer times $2\pi$ ($\pi$)~\cite{supplemental}. Since $k_{z,c}$ scales 
$\sqrt{\delta \beta}$, the conductances show the 
oscillatory behaviors as a function of $\delta \beta L^2_z$. 
Notice also that $G_z$ at the critical point takes a vanishingly small value, 
while the critical conductance value of $G_x$ is much larger. The distinction 
can be attributed to the spatial anisotropy in the clean-limit fixed 
point~\cite{supplemental}.  

\begin{figure}[t]
	\centering
	\includegraphics[width=0.85\linewidth]{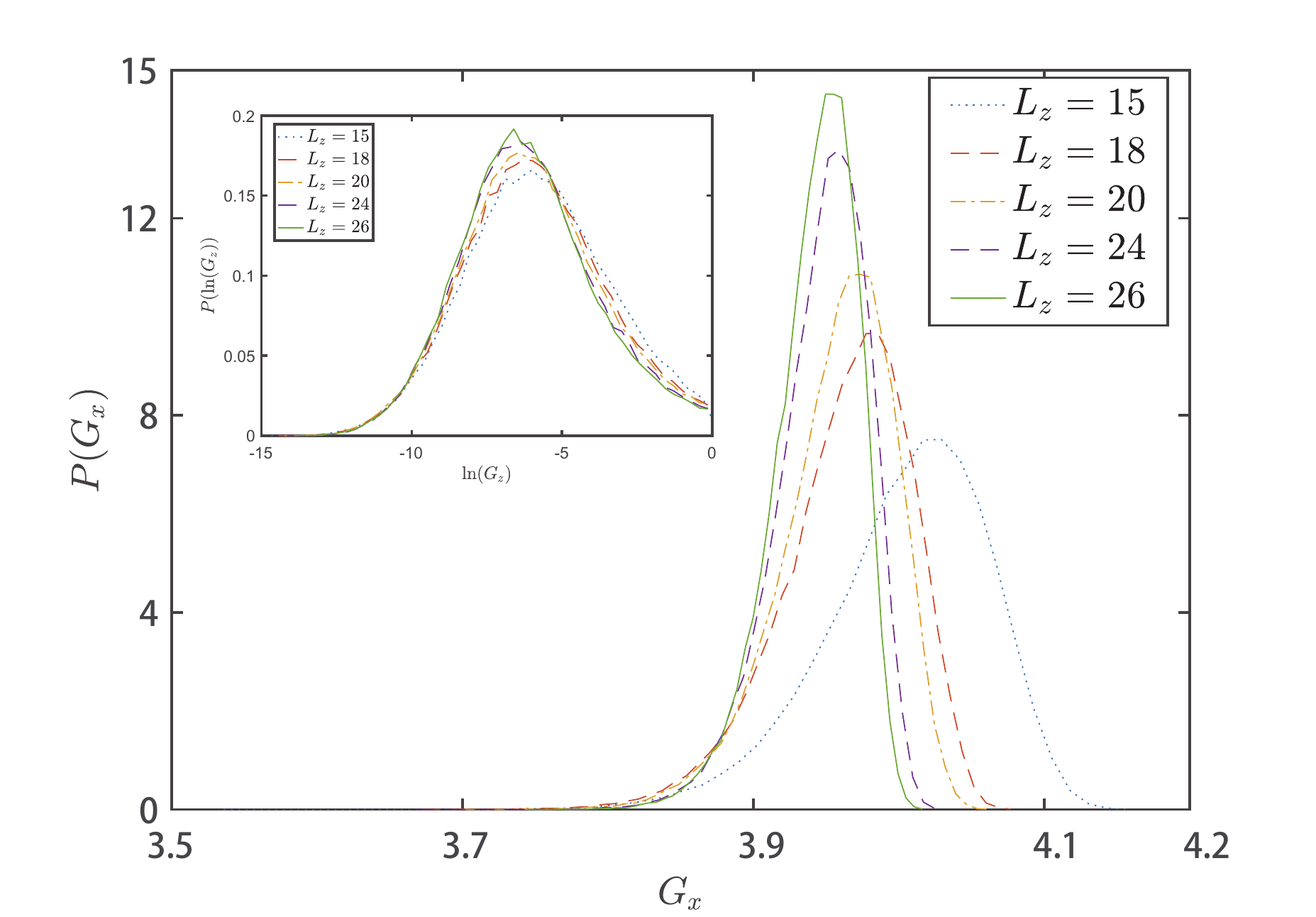}
	\includegraphics[width=0.85\linewidth]{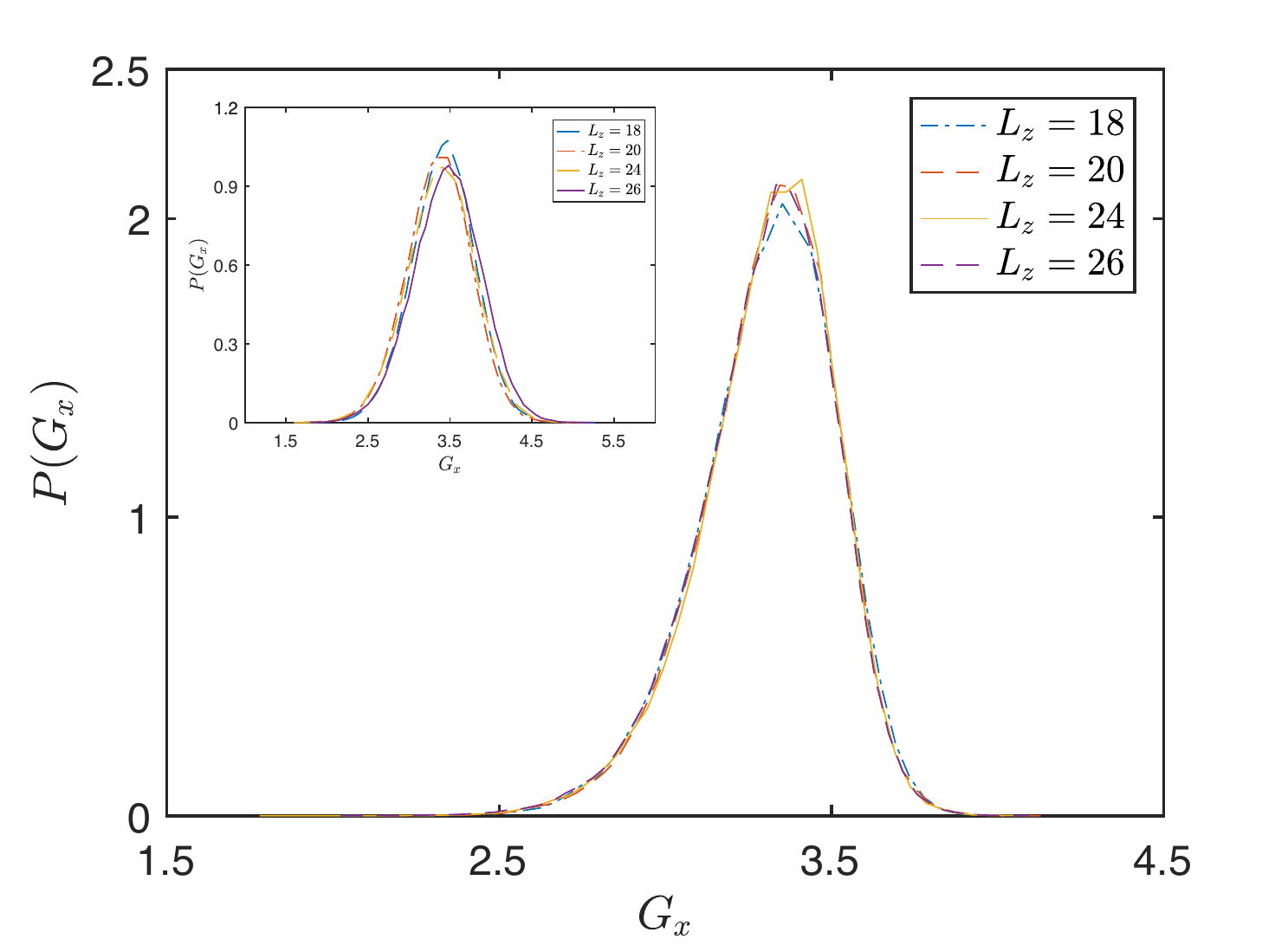}
	\caption{(color online) (upper) Critical conductance distribution 
of $G_x$ (inset: that of $G_z$) on the BI-WSM phase boundary. 
We take a set of parameters in the tight-binding model~\cite{supplemental} 
as $(W,\beta)=(1,0.271)$ (lower) Critical conductance distributions of $G_x$ on 
(or in proximity to)  the critical point (`FP1' in Fig.~\ref{fig:1}) (main) $(W,\beta)=(2.1,0.338)$, 
(inset)  $(W,\beta)=(2.45,0.366)$. We use the tetragonal geometry, 
$L_{\perp}=\eta L^2_z$ with $\eta=1/25$, $(L_z,L_\perp)=(15,9)$, $(18,13)$, $(20,16)$, 
$(24,23)$, $(26,27)$.}
	\label{fig:5}
\end{figure}


The critical conductance distribution (CCD) on the BI-WSM phase boundary 
also shows an unusual scaling behaviour, when 
compared to that of conventional disorder-driven quantum 
phase transitions~\cite{slevin97,slevin00,slevin01prl,slevin01prb,kobayashi10,b-xu16}. 
To see this, let us begin with a general scaling relation between two distribution functions 
of critical conductance, before and after the renormalization;
\begin{align}
P^{\prime}(G_{\mu},L^{\prime}_{\perp},L^{\prime}_z,\Delta^{\prime},m^{\prime}) 
= P(G_{\mu},L_{\perp},L_{z},\Delta,m). \label{CCD-scaling}
\end{align}   
Suppose that the criticality of a quantum phase transition is controlled by 
a fixed point with {\it finite} disorder strength ($\Delta=\Delta_c\ne 0$) and 
with the isotropic scaling ($L^{\prime}_{\perp}=bL_{\perp}$ and $L^{\prime}_z=bL_z$). 
CCD in such conventional quantum phase transition 
point depends only on a system geometry ($L_{\perp}/L_z$) and on universal 
properties encoded in the fixed point. Namely, after a certain times of the renormalization, 
$\Delta^{\prime}$ and others parameters already get (close) to 
a set of values at the fixed point, while 
$L^{\prime}_{\perp}$ and $L^{\prime}_z$ remain much larger 
than the lattice constant scale. Thereby, the right hand side of 
Eq.~(\ref{CCD-scaling}) is essentially equal to the left hand side with 
$\Delta^{\prime} = \Delta_c \ne 0$, 
where the ratio $L^{\prime}_{\perp}/L^{\prime}_z = L_{\perp}/L_z$ 
determines the CCD. 

Our numerical simulation (upper panel of Fig.~\ref{fig:5}) indicates 
that CCD on the BI-WSM phase boundary essentially takes 
a form of the delta function, but the distribution becomes larger for smaller system. 
This is anticipated because the criticality of the phase transition is 
controlled by a fixed point (FP0) with {\it zero} disorder strength ($\Delta_c=0$). Besides,  
the renormalization needs to be truncated when either $L^{\prime}_{\perp}$ 
or $L^{\prime}_z$ reaches the lattice constant scale in the l.h.s. of Eq.~(\ref{CCD-scaling}). 
The truncation results in larger renormalized disorder $\Delta^{\prime}$ 
for smaller $L_{\perp}$ and $L_{z}$.

The BI-WSM phase transition line has an end 
point at a finite disorder strength $\Delta_c \ne 0$, which we dub `FP1' 
as in Fig.~\ref{fig:1}. The critical end point is another scale-invariant fixed point 
and has two relevant scaling variables, $\delta\Delta\equiv \Delta-\Delta_c$ 
and $m$, and numerous irrelevant scaling variables.  On such a fixed point, 
the disorder strength and the effective mass stay at $\Delta_c$ and $0$ respectively, 
while all the irrelevant scaling variables reduce to zero after the renormalization. 
Thus, the CCD calculated with the tetragonal geometry $L_{\perp}=\eta L^2_z$ 
is expected to be scale invariant for fixed geometric parameter $\eta$. 
To see the CCD scale invariance at the critical end point, we calculate the conductance 
distribution for a number of different disorder strength $W$ {\it along} the BI-WSM 
phase transition line. The BI-WSM phase transition line can be accurately determined by the 
self-consistent Born calculation. For a certain disorder strength along 
the BI-WSM boundary line, our numerical results indeed 
observe the CCD scale-invariant feature 
(lower panel of Fig.~\ref{fig:5}) as well as prominent kink-like features in the critical 
conductances $G_x$ and $G_z$~\cite{supplemental}.

\begin{figure}[t]
	\centering
	\includegraphics[width=1.0\linewidth]{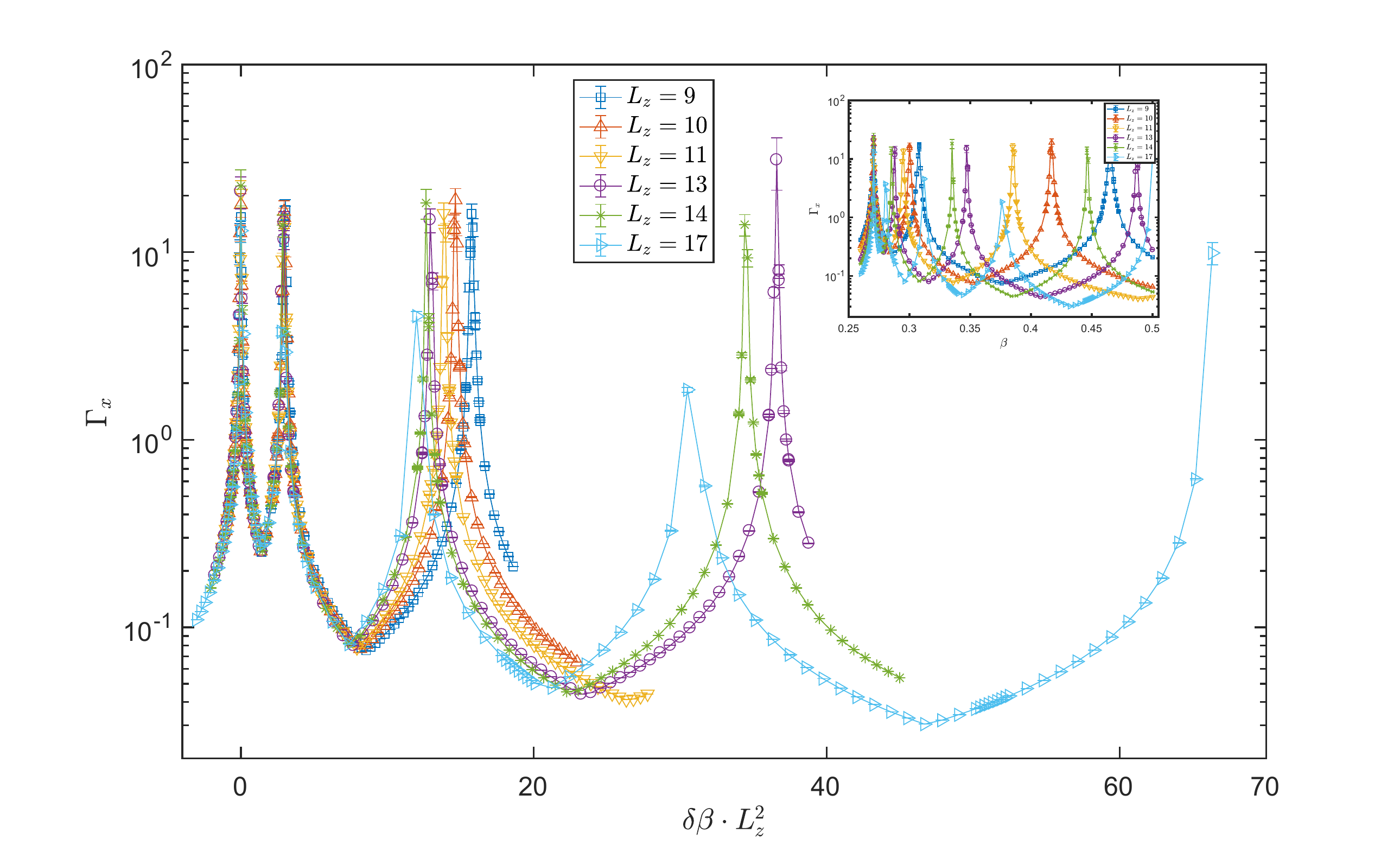}
	\caption{(color online) 
$\Gamma_x\equiv\xi_x/L_z^2$ as a function of $\delta \beta L_z^{2}$ near 
the BI-WSM phase transition point with $\Delta < \Delta_c$. We take a set of 
parameters in the tight-binding model~\cite{supplemental} as 
$(W,\beta)=(1, \beta_c+\delta \beta)$ with $\beta_c=0.2705$ and $\eta=1/10$. 
$\delta \beta$ is proportional to the effective mass $m$ 
defined in Eqs.~(\ref{eff0}) and (\ref{xix-scaling-3}). 
Inset: $\Gamma_x$ as a function of 
$\delta \beta$ with different $L_z$. An oscillatory behaviour of $\xi_x$ as a function of 
$\delta \beta$ is of the same origin as that of the conductance in Fig.~\ref{fig:2+3} (see 
the text).} 
	\label{fig:8}
\end{figure}

The zero-energy localization lengths at the BI-WSM phase transition also obey 
unconventional scaling function forms. From Eq.~(\ref{scaling-zxy}) at $E=0$, 
we obtain RG scaling relations of the localization length along the dipole ($z$) direction 
$\xi_z$ and that along the perpendicular ($xy$) direction $\xi_x$,  
\begin{align}
\xi^{\prime}_x(L^{\prime}_x,L^{\prime}_{y},L^{\prime}_z,\Delta^{\prime},m^{\prime}) 
= b \xi_{x}(L_x,L_y,L_z,\Delta,m), \label{xix-scaling} \\ 
\xi^{\prime}_z(L^{\prime}_x,L^{\prime}_{y},L^{\prime}_z,\Delta^{\prime},m^{\prime}) 
= b^{\frac{1}{2}} \xi_{z}(L_x,L_y,L_z,\Delta,m). \label{xiz-scaling} 
\end{align}
For $\Delta \le \Delta_c$, $L^{\prime}_{x,y}=bL_{x,y}$, $L^{\prime}_z=b^{\frac{1}{2}}L_z$, 
$\Delta^{\prime}=b^{|y_{\Delta}|}\Delta$, $m^{\prime}=b^{-1}m$ with $b<1$. Henceforth, 
we omit dependences of the irrelevant parameter $\Delta$. The RG scaling relations lead 
to the following scaling forms of $\xi_x$ and $\xi_z$; 
\begin{align}
\xi_{x}(L_x,L_y,L_z,m) &= m^{-1} \Psi_x(mL_x,mL_y,m^{\frac{1}{2}}L_z),  \label{xix-scaling-2} \\ 
\xi_{z}(L_x,L_y,L_z,m) &= m^{-\frac{1}{2}} \Psi_z(mL_x,mL_y,m^{\frac{1}{2}}L_z). \label{xiz-scaling-2} 
\end{align}
Eq.~(\ref{xiz-scaling-2}) gives a single-parameter scaling form for 
the quasi one-dimensional system ($L_{x,y}=L_{\perp}\ll L_z$)~\cite{luo18};
\begin{align}
\Gamma_z(L_x=L_y=L_{\perp},m) \equiv \xi_z/\sqrt{L_{\perp}} = \psi_z(mL_{\perp}).  
\label{xiz-scaling-3} 
\end{align}
To verify the scaling form of Eq.~(\ref{xix-scaling-2}) in terms of the transfer matrix 
calculation~\cite{mackinnon81,pichard81}, we set 
$L_y=\eta L^2_z$ and calculate $\xi_x$ for very large 
$L_x = 10^5 \sim 10^7$. For such a geometry, the 
localization length normalized by $L^2_z$ should show scaling invariance  
at the BI-WSM phase transition point ($m=0$);
\begin{align}
\Gamma_x(L_y = \eta L^2_z,L_z,m) 
\equiv \xi_x/L^2_z = \psi_x(m^{\frac{1}{2}}L_z;\eta). \label{xix-scaling-3}
\end{align}
To test this single parameter scaling form in the numerics, we 
take $\eta=1/10$, $(L_z,L_y)=(9,8)$, $(10,10)$, 
$(11,12)$, $(13,17)$, $(14,20)$, $(17,29)$, that approximately satisfy   
$L_y=\eta L^2_z$. Fig.~\ref{fig:8} demonstrates 
that the $\Gamma_x$ with different 
$L_z$ near the phase transition point (small $\delta \beta L_z^2$ region) collapse into a single 
scaling function of $m L^2_z$. 

In this paper, we clarified novel scaling theories of conductance, CCD 
and localization length in the quantum phase transition 
between three-dimensional BI and WSM phases. 
The idea in this paper can be also applied to a direct phase transition between 
ordinary band insulator (OI) and 
topological insulator (TI) phases~\cite{shindou09prb,kobayashi13,kobayashi14}, 
whose criticality is controlled by a clean-limit fixed point. The conductance 
scaling at the OI-TI phase transition is given by Eq.~(\ref{conventional}) with 
$\nu=1$, while the CCD on the boundary line takes a delta function form. 

A recent transport experiment discovered a solid-state material that exhibits continuous 
BI-WSM phase transitions~\cite{liang17}. Our paper reveals the universal critical properties 
of this continuous phase transition through the electric conductance. The results show 
that the conductance at the critical point is scaled by $L^2_z/L_{\perp}$: conventionally,  
the critical conductance is scaled by $L_z/L_{\perp}$. Such difference in the conductance scaling 
has significant impact on the transport experiment, compared to mere differences 
in the critical exponent.

\begin{acknowledgements}
This work (X. L., and R. S.) was supported by NBRP of China Grants No.~2014CB920901, 
No.~2015CB921104, and No.~2017A040215. T. O. was supported by JSPS KAKENHI 
Grants No.~JP15H03700 and No.~JP17K18763.  
\end{acknowledgements}

\clearpage
\begin{widetext}
\begin{center}
\textbf{Supplemental Material}
\end{center}

\section{tight-binding model for a layered Chern insulator with disorder}
In order to test the new scaling functions numerically, we use 
a two-orbital tight-binding model defined on a cubic lattice~\cite{luo18,cz-chen15,s-liu16}. 
The model consists of an $s$ orbital and a $p= p_{x}+ip_{y}$ orbital on each cubic 
lattice site (${\bm x}$);  
\begin{align}
{\cal H} = & \sum_{{\bm x}} \Big\{
\big(\epsilon_s + v_s({\bm x})\big) c^{\dagger}_{{\bm x},s} c_{{\bm x},s}
+ \big(\epsilon_p + v_p({\bm x})\big) c^{\dagger}_{{\bm x},p} c_{{\bm x},p} \Big\}  
+ \sum_{{\bm x}} \Big\{ - \sum_{\mu=x,y} \big(
t_s c^{\dagger}_{{\bm x} + {\bm e}_{\mu},s} c_{{\bm x},s}
- t_p c^{\dagger}_{{\bm x} + {\bm e}_{\mu},p} c_{{\bm x},p}\big)  \nonumber \\
& +  t_{sp} (c^{\dagger}_{{\bm x}+{\bm e}_x,p}
- c^{\dagger}_{{\bm x} - {\bm e}_x,p}  -  i c^{\dagger}_{{\bm x}+{\bm e}_y,p}
+ i c^{\dagger}_{{\bm x} - {\bm e}_y,p})  \!\ c_{{\bm x},s} 
-t^{\prime}_s c^{\dagger}_{{\bm x} + {\bm e}_{z},s} c_{{\bm x},s}
- t^{\prime}_p c^{\dagger}_{{\bm x} + {\bm e}_{z},p} c_{{\bm x},p} + {\rm H.c.} \Big\}. \label{tb1}
\end{align}
$\epsilon_{s}$, $\epsilon_{p}$ and $\upsilon_{s}(\bm{x})$, $\upsilon_{p}(\bm{x})$, 
are atomic energies for the $s$, $p$ orbital and on-site disorder potential of 
the $s$, $p$ orbital, respectively. The disorder potentials are uniformly 
distributed within $[-W/2,W/2]$ with identical probability distribution. 
$t_{s}$, $t_{p}$, and $t_{sp}$ are intralayer transfer integrals between $s$ orbitals of 
nearest neighboring two sites, that between $p$ orbitals, and that 
between $s$ and $p$ orbitals, respectively, while $t_{s}^{\prime}$ and $t_{p}^{\prime}$ are 
interlayer transfer integrals. ${\bm e}_{\mu}$ ($\mu=x,y$) are primitive translational vectors 
within a square-lattice plane. ${\bm e}_{z}$ is a primitive translational vector 
connecting neighboring square-lattice layers.  In this paper, we take 
\begin{align}
t_p &= t_s, \ \ \epsilon_s = -\epsilon_p = 2t_s + 4t_s \beta, \nonumber \\ 
t^{\prime}_p &= - t^{\prime}_s = 2t_s \beta, \ \ t_{sp} = \frac{4t_s}{3}, \nonumber 
\end{align}
where we change an `interlayer coupling strength' $\beta$. The tight-binding 
model without disorder ($W=0$) reduces to the following 2 by 2 Hamiltonian, 
\begin{eqnarray}
H({\bm k}) = {\bm a}({\bm k})\cdot {\bm \sigma}. \nonumber 
\end{eqnarray}
Here ${\bm \sigma}=(\sigma_x,\sigma_y,\sigma_z)$ are Pauli matrices and 
\begin{align}
a_{z}({\bm k}) &\equiv 2t_s (1+2\beta) - 2t_s (\cos k_x a_{\perp} + \cos k_y a_{\perp}) + 
4t_s \beta \cos k_z a_z, \label{a3} \\
a_{x}({\bm k}) &\equiv -\frac{8t_s}{3} \sin k_y a_{\perp}, 
\ \ a_{y}({\bm k}) \equiv -\frac{8t_s}{3} \sin k_x a_{\perp}. \label{a12}  
\end{align}
For later clarity, we made explicit lattice constants of the cubic model along the $z$-direction $a_z$ and 
along the $xy$ directions $a_{\perp}$. 

For $\beta<1/4$, the electronic system at the half filling ($E=0$)
is in a 3-dimensional band insulator phase. For $\beta=1/4$, a pair of Weyl nodes is created 
at $E=0$ at the $\Gamma$ point. For $\beta>1/4$, the Weyl 
nodes appear at ${\bm k}=(0,0,\pm k_{z,c})$ with $\cos k_{z,c} = 1/(2\beta)-1$. This 
gives out the clean-limit electronic phase diagram as in the inset of Fig.~1 in the main 
text. Throughout the paper, the unit of the disorder strength $W$ is taken to be $4t_s$.   

\section{conductance oscillations in Weyl semimetal phases}

The two-terminal conductances in the WSM phase side 
show oscillatory behaviors as a 
function of $\delta \beta L^2_z$ (Fig.~2 in the main text). 
For positive $\delta \beta$ (WSM phase side), the Weyl points appear at 
${\bm k}_{\rm MM}=(0,0,\pm k_{z,c})$. In the following, we show that the oscillation behaviour 
of $G_z$ (lower panel of Fig.~2 in the main text) comes from a `commensurability effect' 
between $\pi/L_z$ and the Weyl node position along the $k_z$-axis, $k_{z,c}$.  

In the transfer-matrix calculation of the two-terminal conductance $G_z$, 
two lead Hamiltonians ($z<0$ and $z>L_z$) are attached to the bulk Hamiltonian ($0<z<L_z$). 
The lead Hamiltonian comprises of $L_{\perp}\times L_{\perp}$ number of decoupled 
one-dimensional chains,
\begin{align}
{\cal H}_{\rm lead} = - t_0 \sum_{\bm x} \sum_{a=s,p} 
\Big(c^{\dagger}_{{\bm x}+{\bm e}_z,a} c_{{\bm x},a} + {\rm H.c.}\Big). \label{lead}
\end{align} 
This has $2 \times L^2_{\perp}$ number of the zero-energy eigenstates with positive 
velocity along the $z$ direction. 
The zero-energy conductance is calculated from transmission coefficient between 
the zero-energy eigenstates in the $z<0$ region and those in the $z>L_z$ region. 
Contrary to the large 
number of the zero-energy eigenstates in the two leads, the bulk has 
less than four zero-energy eigenstates. Thereby, most of the zero-energy states 
injected from the lead regions are reflected backward at the two contacts. 
For the bulk wavefunction's point of view, this situation can be effectively described 
by the fixed boundary condition imposed at $z=0$ and $z=L_z$. 

The finite-size system with the fixed boundary condition sees the two Weyl nodes 
only when $k_{z,c}$ becomes equal to $\pi/L_z$ times integer. Thus, the zero-energy 
conductance along the $z$ direction is expected to show a peak structure when 
$k_{z,c}\cdot L_z$ becomes integer times $\pi$. In Fig.~\ref{sfig:4}, we replot 
the data points of $G_z$ in Fig.~2 in the main text as a function of $k_{z,c}\cdot L_z$. 
Here $k_{z,c}$ in the presence of finite disorder strength is accurately determined 
by the self-consistent Born calculation. In Fig.~\ref{sfig:4}, one can see that peak 
structures appear when $k_{z,c}\cdot L_z = n\pi$ ($n=2,3,4,5,\cdots$).  

\begin{figure}[t]
	\centering
	\includegraphics[width=0.8\linewidth]{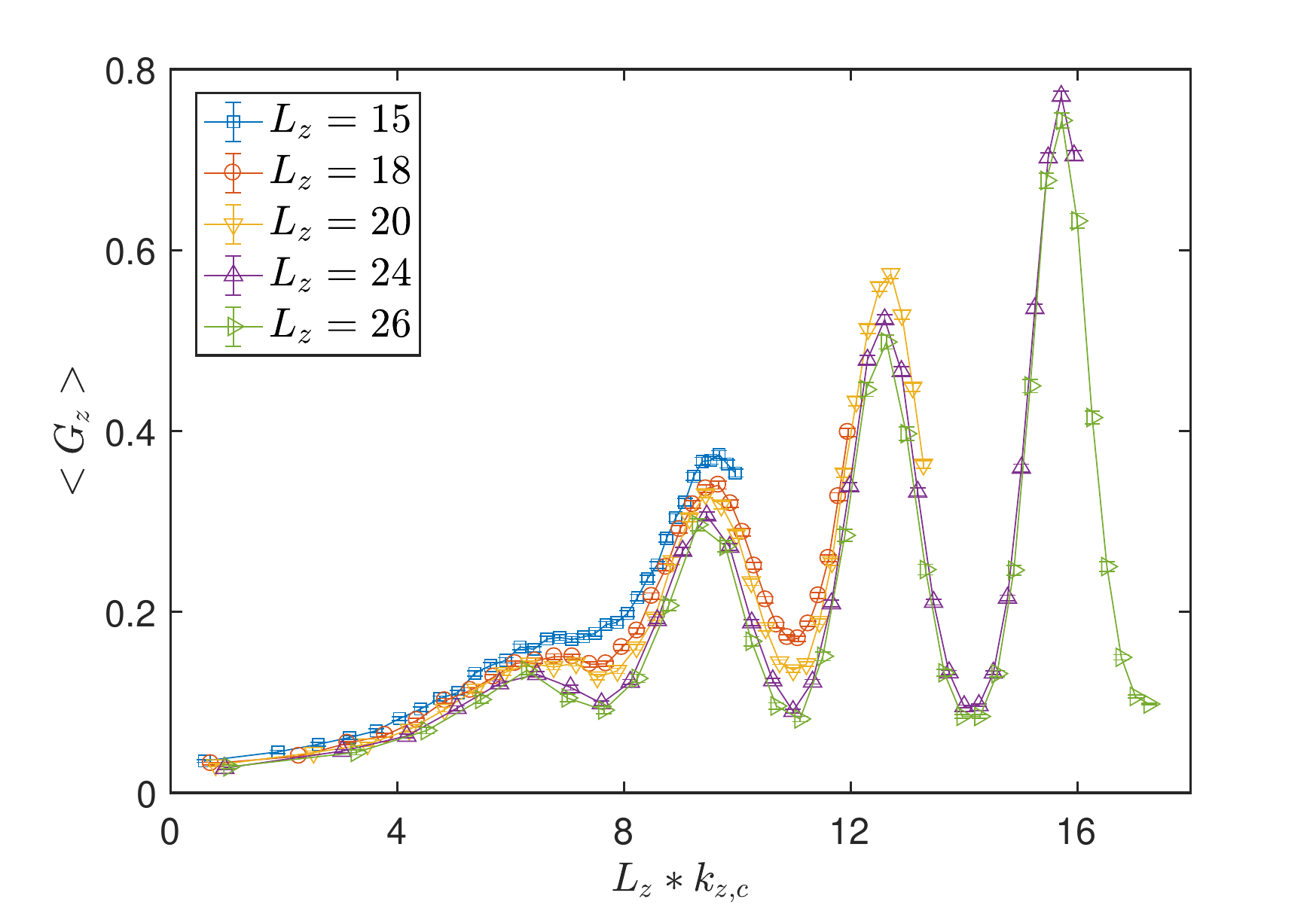}
	\caption{(color online) $G_z$ as a function of $k_{z,c} L_z$ 
reproduced from the same data stream as in Fig.~2 in the main text. 
$k_{z,c}$ in the presence of finite disorder strength 
is determined by the self-consistent Born calculation. 
The interval between neighboring peaks in the plot is around 
$\pi$.}
	\label{sfig:4}
\end{figure}
  
\section{marginal scaling variables and critical conductance}
$b_2$ and $v$ in the effective continuum model, Eq.~(1) in the main text, do not change 
under the renormalization. In other words, `FP0' with 
different $b_2$ and $v$ comprise a surface of fixed points (`fixed surface') 
in a higher-dimensional parameter space that includes these two 
marginal scaling variables (see Fig.~\ref{sfig:1}). The scaling functions of the zero-energy conductance 
and localization length depend on these two scaling variables as; 
\begin{align}
G_{\mu}(L_{\perp},L_z,m,b_2,v,\Delta,\cdots) &= \Phi_{\mu}(mL_{\perp},m^{\frac{1}{2}}L_z,b_2,v,
m^{-y_{\Delta}}\Delta,\cdots), \label{A} \\
\xi_{x}(L_x,L_y,L_z,m,b_2,v,\Delta,\cdots) & = m^{-1} \Psi_{x}(mL_x,mL_y,m^{\frac{1}{2}}L_z,
b_2,v,m^{-y_{\Delta}}\Delta,\cdots). \label{B}
\end{align}
From Eq.~(\ref{a3}), one can see that the interlayer coupling strength $\beta$ changes 
the effective mass $m$ as well as one of the two marginal parameters, $b_2=2t_s \beta a^2_z$ 
[$v$ does not depend on $\beta$; see Eq.~(\ref{a12})]. 
Thereby, the $b_2$-dependence in Eqs.~(\ref{A},\ref{B}) 
can result in larger deviations of the data points from the single parameter 
scaling forms in Figs~2, 4 in the main text. 

\begin{figure}[t]
	\centering
	\includegraphics[width=0.75\linewidth]{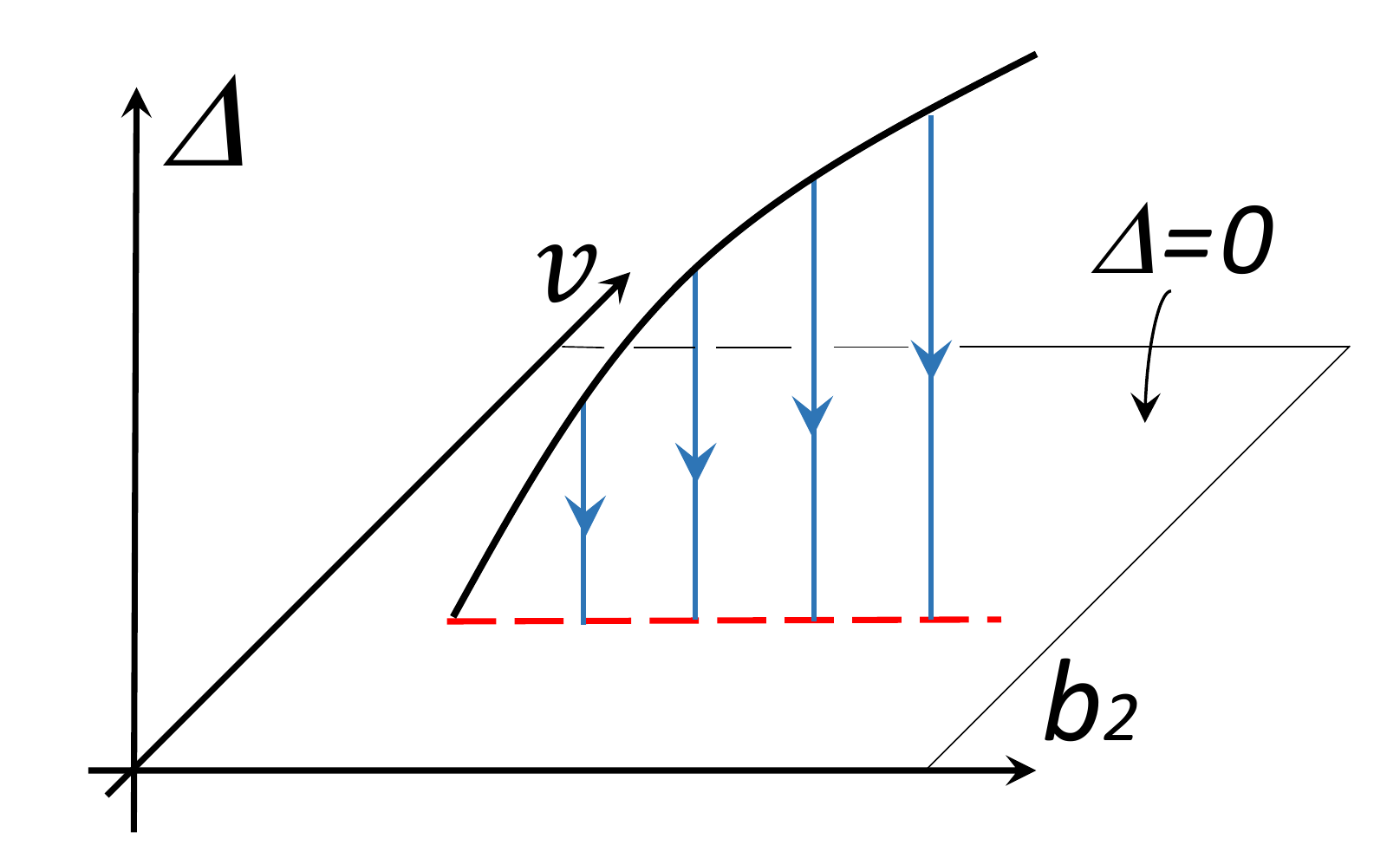}
	\caption{(color online) Schematic figure that explains how the BI-WSM phase boundary 
line in the inset of Fig.~1 in the main text is seen in the three-dimensional 
parameter space subtended by the disorder strength $\Delta$, 
and two marginal parameters $b_2$ and $v$. The $\Delta=0$ 
plane in the figure is a fixed surface, within which any points do not move under the renormalization. 
The blue arrows denote RG flows, connecting the BI-WSM phase boundary line with finite 
disorders (black solid line) with its projection onto the fixed surface (red dotted line).}
	\label{sfig:1}
\end{figure}

The scale-invariant value of critical conductance $G_x$ varies with the disorder strength 
along the BI-WSM phase boundary line (see Fig.~\ref{sfig:7}). This is apparently inconsistent 
with the $\Delta$-dependence proposed in Eqs.~(5,6) in the main text. The discrepancy can be resolved, 
once we take into account $b_2$ and $v$-dependences as in Eqs.~(\ref{A},\ref{B}). Namely, 
under the renormalization, the critical conductance value on the BI-WSM boundary line 
with finite disorder $\Delta$ can be equated to that on its projection line at $\Delta=0$ 
(see Fig.~\ref{sfig:1});
\begin{align}
\lim_{L_z \rightarrow +\infty} G_{\mu}(L_{\perp}=\eta L^2_z,L_z,m=0,b_2,v,\Delta,\cdots) 
= \lim_{L^{\prime}_z \rightarrow +\infty} G_{\mu}(L^{\prime}_{\perp}=
\eta {L^{\prime}_z}^2,L^{\prime}_z,m=0,b_2,v,\Delta=0,\cdots). \label{C}
\end{align} 
Here $b_2$ in Eq.~(\ref{C}) is given by the critical interlayer coupling strength $\beta_c$; 
Eq.~(\ref{a3}) gives $b_2=2t_s \!\ \beta_c \!\ a^2_z$. Meanwhile, the critical interlayer coupling strength 
changes with the disorder strength (see the inset of Fig.~1 in the main text). 
Thus, the left hand side of Eq.~(\ref{C}) can vary with the disorder strength 
$\Delta$ through the $b_2$-dependence of the right hand side with $b_2=2t_s \!\ \beta_c (\Delta) \!\ a^2_z$. 

The critical conductance value of $G_{\mu}$ is given as a function of the two 
marginal scaling variables as well as the geometric parameter $\eta$ with $L_{\perp}=\eta L^2_z$. 
From the dimensional analysis, the scaling form is given by 
\begin{align}
\lim_{L_z \rightarrow +\infty} G_{\mu}(L_{\perp}=\eta L^2_z,L_z,m=0,b_2,v,\Delta=0,\cdots) 
= \Theta_{\mu}\Big(\frac{b_2 \eta a_{\perp}}{v a^2_{z}}\Big), \label{D}
\end{align}
where $a_{z}$ and $a_{\perp}$ are lattice constants of the tight-binding model along the dipole ($z$)
direction and the perpendicular ($xy$) directions respectively. 
The scaling function $\Theta_{\mu}(x)$ can be evaluated explicitly. For the zero-energy 
conductance along the perpendicular directions ($\mu=x,y$), the function is given by 
\begin{align}
\Theta_{x}\Big(\frac{b_2\eta a_{\perp}}{v a^2_{z}}\Big)
= \lim_{L_z \rightarrow +\infty} \lim_{E \rightarrow +0} \frac{e^2}{\hbar} \sum_{j_y}\sum_{j_z} 
\int^{\infty}_{0} \frac{dk_x}{2\pi} \frac{\partial E_{\bm k}}{\partial k_x} \delta(E - E_{\bm k}), \label{E}
\end{align}
where 
\begin{align}
E_{\bm k} &\equiv \sqrt{b^2_2k^4_z + v^2(k^2_x+k^2_y)}, \label{E1} \\
k_y & = \frac{2\pi}{L_{\perp}a_{\perp}} j_y,  \!\ k_z = \frac{2\pi}{L_za_{z}} j_z, \!\ L_{\perp} = \eta L^2_z, \label{E2} 
\end{align}
with two integers $j_y,j_z=0,\pm 1,\pm 2, \cdots$. Note that there exists a subtlety in 
the order of the two limits in Eqs.~(\ref{E}), $\lim_{E \rightarrow +0}$ and 
$\lim_{L_z\rightarrow +\infty}$. When we take $E$ to be zero first and then take $L_z$ to be 
infinite, Eq.~(\ref{E}) reduces to $e^2/h$ times a number of those zero-energy eigenstates 
that carry positive velocities along the $x$ direction. The quantum critical point of the magnetic 
dipole model ($\Delta=m=0$) has only one such zero-energy eigenstate. 
The quantization in $e^2/h$ is obviously inconsistent with the numerical 
observations in Fig.~2 in the main text and Fig.~\ref{sfig:7}. The discrepancy can be resolved, once we 
reconsider the order of the two limits.  

In the transfer-matrix calculation of the two-terminal conductance, 
a whole system consists of the bulk Hamiltonian and two lead Hamiltonians 
that are attached to the bulk. For the lead Hamiltonian, 
we employ a `flux' model as in Eq.~(\ref{lead}).  The zero-energy conductance 
is calculated from the transmission coefficient between the zero-energy eigenstates of 
the lead Hamiltonian in the $z<0$ region and those in the $z>L_z$ region. 
Such zero-energy eigenstates are {\it not} eigenstates of the bulk Hamiltonian. Instead,  
they are superpositions of those eigenstates of the bulk Hamiltonian that distribute around $E=0$. 
For such cases, it is more natural taking the two limits simultaneously rather than 
taking $E$ to be strictly zero from the outset. 
For a finite-size system with the tetragonal geometry, the eigen energy 
near $E=0$ is discretized by either $\frac{4\pi^2 \!\ b_2}{(L_za_{z})^2}$ or 
$\frac{2\pi \!\ v}{L_{\perp} a_{\perp}}$. We thus set the single particle energy 
$E$ in Eq.~(\ref{E}) to be $\frac{2\pi \!\ v\gamma}{L_{\perp}a_{\perp}}$. Here small 
$\gamma$ represents  a dimensionless quantity that quantifies  
nature of the contact between lead and bulk. 
Generally, $\gamma$ is smaller for a better contact. 

\begin{figure}[t]
	\centering
	\includegraphics[width=0.9\linewidth]{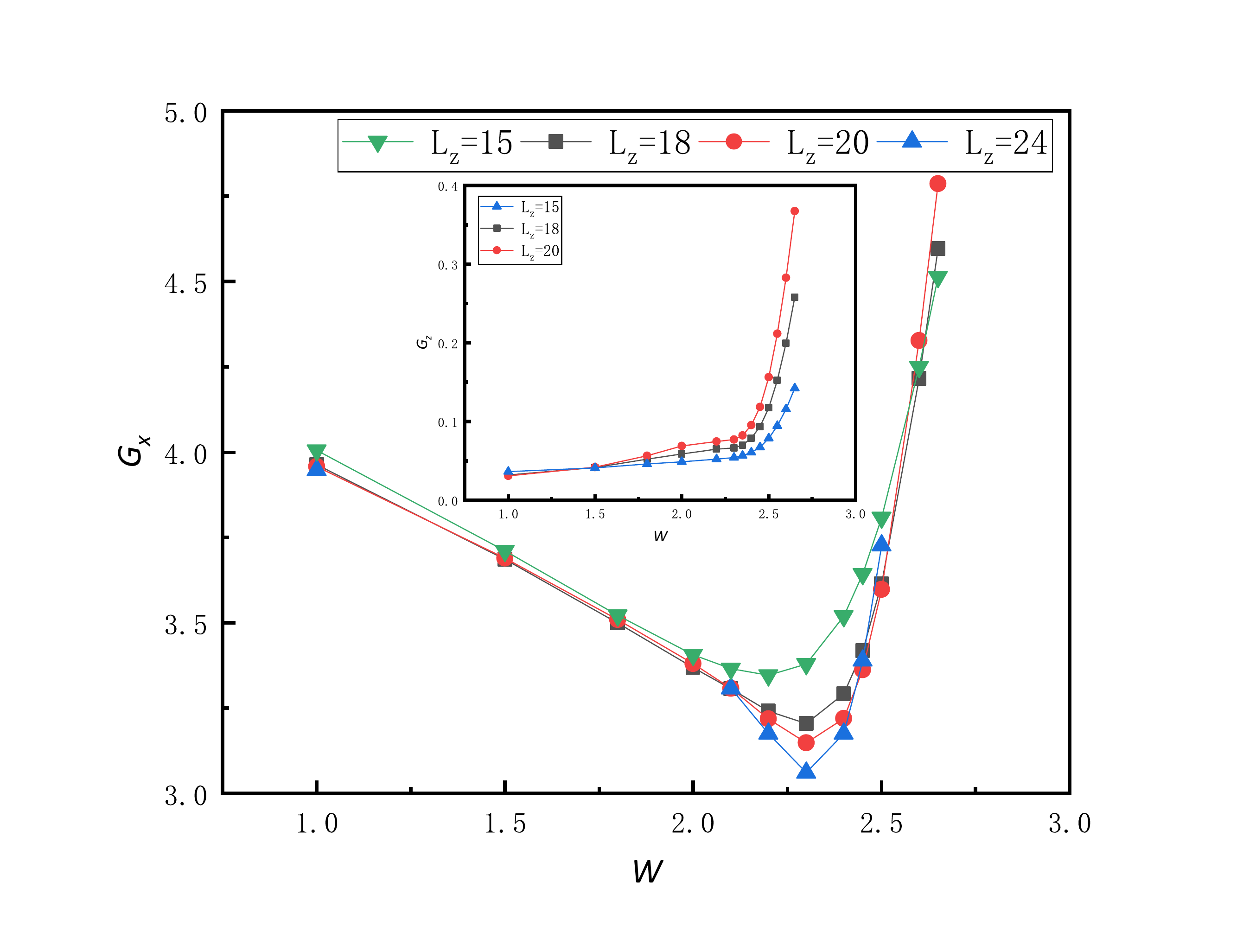}
	\caption{(color online) Critical conductance value of 
$G_x$ (inset: that of $G_z$) with $10^5$ samples as a 
function of the disorder strength $W$ {\it along} the BI-WSM phase transition line. 
The BI-WSM phase transition line is accurately determined by the self-consistent Born 
calculation.}
	\label{sfig:7}
\end{figure}

The critical conductance along the $x$ direction is 
calculated in this intermediate limit as, 
\begin{align}
\Theta_x\Big(\frac{b_2\eta a_{\perp}}{v a^2_z}\Big) &= \lim_{L_{z} \rightarrow +\infty} 
 \frac{e^2}{\hbar} \sum_{j_y}\sum_{j_z} 
\int^{\infty}_{0} \frac{dk_x}{2\pi} \frac{\partial E_{\bm k}}{\partial k_x} 
\delta(E - E_{\bm k})_{|E=\frac{2\pi v\gamma}{L_\perp a_{\perp}}}  \nonumber \\
&= \frac{e^2}{h} 
\sqrt{\frac{v \gamma^3 a^2_{z}}{2\pi b_2 \eta a_{\perp}}}  
\frac{\sqrt{\pi} \!\ \Gamma\big(\frac{1}{4}\big)}{2\Gamma\big(\frac{7}{4}\big)} 
= \frac{e^2}{h} 
\sqrt{\frac{v \gamma^3 a^2_{z}}{2\pi b_2 \eta a_{\perp}}} \times 3.49\cdots, \label{F}  
\end{align}
where $\Gamma(x)$ is the Gamma function. The critical 
conductance along the dipole ($z$) direction 
is calculated in the same limit with $E=\frac{2\pi v\delta}{L_{\perp} a_{\perp}}$, 
\begin{align}
\Theta_{z}\Big(\frac{b_2\eta a_{\perp}}{v a^2_z}\Big)= \lim_{L_z \rightarrow +\infty} 
\frac{e^2}{\hbar} \sum_{j_x}\sum_{j_y} 
\int^{\infty}_{0} \frac{dk_z}{2\pi} \frac{\partial E_{\bm k}}{\partial k_z} 
\delta(E - E_{\bm k})_{|E=\frac{2\pi v\delta}{L_\perp a_{\perp}}} 
= \frac{e^2}{h}  \pi \delta^2.  \label{G}  
\end{align}
Here $\gamma$ and $\delta$ are generally different from each other. Namely, $\gamma$ 
characterizes the contact between the lead and the $yz$ plane of the bulk Hamiltonian 
Eq.~(\ref{tb1}), while $\delta$ characterizes a contact between the lead with th $xy$ plane 
of Eq.~(\ref{tb1}). For the tight-binding model described above, $b_2=2t_s\beta a^2_z$, 
$v= \frac{8t_s}{3}a_{\perp}$, and $\eta=1/25$. This gives $(b_2\eta a_{\perp})/(va^2_{z})=0.81/100$ 
at $\beta=\beta_c=0.27\cdots$.  For $\gamma=0.4$ and 
$\delta=0.1$, we have 
\begin{align}
G_{x} = \frac{e^2}{h} \times 3.91 \cdots, \ \ \ 
G_{z} = \frac{e^2}{h} \times 0.0314\cdots .  
\end{align}
These values are consistent with the order of the critical conductance values shown in 
Fig.~2 in the main text. Notice also that Eq.~(\ref{G}) has no explicit $b_2$ dependence. 
This is consistent with a weak $W$-dependence of the critical conductance value 
$G_{z}$ along the BI-WSM phase boundary line as in the inset of Fig.~\ref{sfig:7}.

\section{density of states}
According to the preceding theories~\cite{roy16arXiv,luo18},  
the density of states at a single-particle energy 
$E$ follows $\rho(E,\Delta,m)=m^{d-\frac32} \Omega(m^{-1} E)$ 
around the BI-WSM phase transition line ($d=3$). A universal scaling 
function $\Omega(x)$ vanishes quadratically in small $x$ 
region (`magnetic monopole regime'), while it diverges as $x^{\frac{3}{2}}$ 
in large $x$ region (`magnetic dipole regime'). Fig.~\ref{sfig:2} demonstrates 
a crossover between these two different critical regions.  
\begin{figure}[t]
	\centering
	\includegraphics[width=0.75\linewidth]{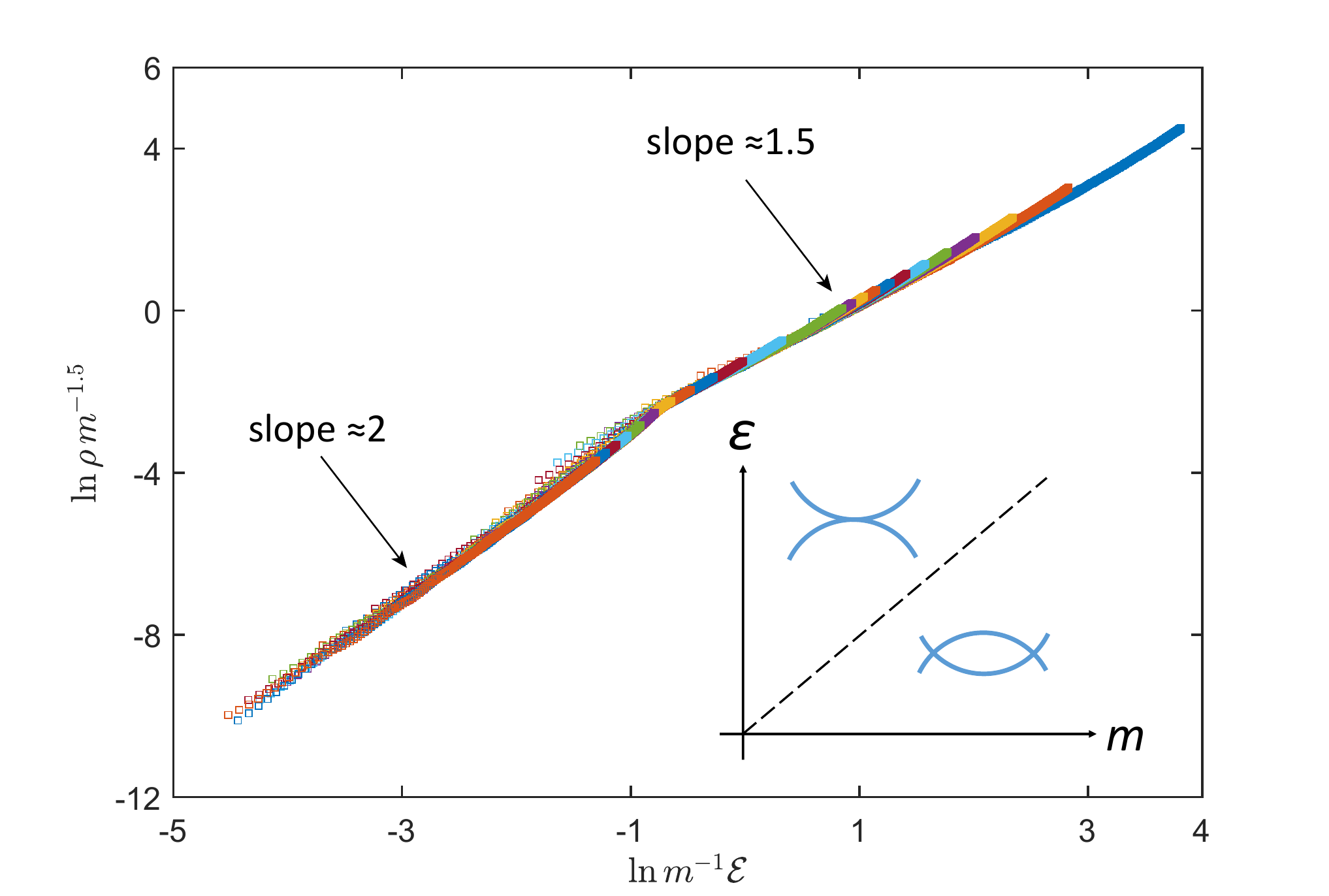}
	\caption{(color online) 
Log-Log plot of $m^{-d+\frac32}\rho(E)$ as a function 
of $m^{-1}E$. We take a set of parameters in the tight-binding model Eq.~(\ref{tb1})   
as $(W,\beta)=(1,0.271+\delta \beta)$ and  
$0<\delta\beta<0.25$. $\delta \beta$ is proportional to the effective mass
 $m$ in Eq.~(1) in the main text. The density of states $\rho(E)$ is calculated 
in terms of the kernel polynomial expansion method. For data with 
$\beta$ larger/smaller than 0.29, we set the Chebyshev expansion 
order $N$ to be $N=2000/1500$.}
	\label{sfig:2}
\end{figure}

\end{widetext}
\end{document}